\documentclass[]{aa}  

\usepackage{graphicx}
\usepackage{txfonts}

\begin{document}

\title{The contribution  of red  dwarfs and white  dwarfs to  the halo
       dark matter}

\author{S. Torres\inst{1,2}, 
        J. Camacho\inst{1},
        J. Isern\inst{3,2} \and 
        E. Garc\'{\i}a--Berro\inst{1,2}}
\institute{Departament de F\'\i sica Aplicada, 
           Escola Polit\'ecnica Superior de Castelldefels, 
           Universitat Polit\`ecnica de Catalunya,
           Avda. del  Canal Ol\'\i mpic 15, 
           08860 Castelldefels, Spain\
           \and       
           Institute for Space Studies of Catalonia,
           c/Gran Capit\`a 2--4, Edif. Nexus 104,   
           08034  Barcelona,  Spain\ 
           \and
           Institut de Ci\`encies de l'Espai, CSIC,  
           Campus UAB, Facultat de Ci\`encies, Torre C-5, 
           08193 Bellaterra, Spain\ }

\offprints{E. Garc\'\i a--Berro}

\date{\today}

\abstract{The nature  of the  several microlensing events  observed by
the MACHO team towards  the LMC still remains controversial. Low--mass
substellar objects and stars with masses larger than $\sim1\,M_{\sun}$
have  been ruled out  as major  components of  a MACHO  galactic halo,
while stars of half solar masses are the most probable candidates.}
{In this  paper we assess  jointly the relative contributions  of both
red dwarfs and white dwarfs to the mass budget of the galactic halo.}
{In  doing  so we  use  a  Monte  Carlo simulator  which  incorporates
up--to--date  evolutionary  sequences of  both  red  dwarfs and  white
dwarfs as well as detailed descriptions  of our Galaxy and of the LMC.
We explore the  complete mass range between $0.08$  and $1\, M_{\sun}$
as  possible  microlensing candidates  and  we  compare the  synthetic
populations obtained  with our simulator with the  results obtained by
the MACHO and EROS experiments.}
{Our results indicate  that, despite that the contribution  of the red
dwarf population increases  by a factor of 2 the  value of the optical
depth  obtained when taking  into account  the white  dwarf population
alone,  it is  still  insufficient  to explain  the  number of  events
claimed by the MACHO team.}   
{Finally, we find that the contribution to the halo dark matter of the
whole  population under  study is  smaller than  $10\%$ at  the $95\%$
conficence level.}


\titlerunning{The contribution  of red dwarfs and white  dwarfs to the
              halo dark matter}

\authorrunning{S. Torres et al. }

\maketitle


\section{Introduction}

Several  cosmological  observations   show  compelling  evidence  that
baryons represent  only a  small fraction of  the total matter  in our
Universe   and   that  non--baryonic   dark   matter  dominates   over
baryons. Within  the standard cosmological model  this naturally leads
to the existence  of a sort of unknown energy,  the dark energy, which
is  the predominant  component,  $\Omega_{\Lambda}\simeq 0.72$,  along
with the  dark matter component, $\Omega_{\rm  M}\simeq 0.27$, whereas
the   baryonic  component  is   just  $\Omega_{\rm   B}\simeq  0.044$.
Moreover,   most    of   the   baryons    are   non--luminous,   since
$\Omega_\star\simeq 0.005$.   For the  case of our  own Galaxy  it has
been found that the virial mass out to 100 kpc is $M\approx 10^{12} \,
M_{\sun}$ while the  baryonic mass in the form  of stars is $M_{\star}
\approx 7 \times  10^{10}\, M_{\sun}$, which means that  for the Milky
Way, the baryon  fraction is at most 8\% (Klypin  et al.  2007).  This
problem is known  as the missing baryon problem ---  see the review of
Silk (2007) for a complete,  interesting and recent discussion of this
issue ---  and it is critical  in our understanding of  how the Galaxy
(an  by extension  other  galaxies) were  formed  and will  ultimately
evolve.  To  solve this problem, three alternatives  can be envisaged:
either  these baryons  are in  the outer  regions of  our  Galaxy, or,
perhaps, they never were present  in the protogalaxy or, finally, they
may  have  been  ejected  from  the Milky  Way.   The  most  promising
explanation  and the  currently  favored  one is  the  first of  these
options.

Since the pioneering proposal of Paczy\'nski (1986) that gravitational
microlensing  could be  used to  clarify the  nature of  galactic dark
matter,  considerable  observational   and  theoretical  efforts  have
invested in this issue.  Among the most likely candidates for building
up the baryonic dark matter density are massive baryonic halo objects,
or MACHOs.  From  the theoretical point of view  it has been suggested
that MACHOs  could be planets ($M\sim 10^{-7}\,  M_{\sun}$), brown and
red   dwarfs  (with  masses   ranging  from   $\sim  0.01$   to  $\sim
1\,M_{\sun}$),  primordial black holes  ($M\,\ga 10^{-16}\,M_{\sun}$),
molecular clumps  ($M\sim1\, M_{\sun}$)  and old white  dwarfs ($M\sim
0.6  \,  M_{\sun}$).  From  the  observational  perspective the  MACHO
(Alcock et al.   1997, 2000), EROS (Lasserre et  al.  2001; Goldman et
al.  2002; Tisserand  et al.  2007), OGLE (Udalski  et al.  1994), MOA
(Muraki et al.  1999) and  SuperMACHO (Becker et al.  2005) teams have
monitored millions  of stars  during several years  in both  the Large
Magellanic Cloud (LMC) and the  Small Magellanic Cloud (SMC) to search
for microlensing  events. However,  only the MACHO  team and  the EROS
group  have  reported  their  results.  Despite  that  initially  some
divergences appeared between their results, agreement has been reached
in some  important points.   For instance, no  microlensing candidates
have  been found  by  the MACHO  team  or the  EROS  group with  event
durations  between  a few  hours  and 20  days.  This  means that  the
Galactic halo could contain no more than a 10\% of dark objects in the
mass range $10^{-7}< M/M_{\sun}<10^{-3}$.   This rules out planets and
brown dwarfs as the major contributors  to the mass budget of the dark
halo.   Moreover the  MACHO collaboration  has succeeded  in revealing
$\sim 15$  microlensing events  during their 5.7  yr analysis  of 11.9
million stars  in the  LMC (Alcock et  al.  2000).  In  their analysis
they    derived    an   optical    depth    towards    the   LMC    of
$\tau=1.2^{+0.4}_{-0.3}\times  10^{-7}$   or,  equivalently,  an  halo
fraction  $0.08<f<0.50$ at the  $95\%$ confidence  level with  a MACHO
mass in  the range $0.15\leq  M/M_{\sun} \leq 0.50$, depending  on the
halo model. On the other hand,  the negative results found by the EROS
collaboration provide an upper  limit. Specifically, the EROS team has
presented an analysis of a  subsample of bright stars belonging to the
LMC  (Tisserand  et  al.   2007),  in order  to  minimize  the  source
confunsion and blending effects.  Their results imply that the optical
depth  towards the  LMC  is $\tau<0.36\times  10^{-7}$  at the  $95\%$
confidence level,  corresponding to  a fraction of  halo mass  of less
than $7\%$. This  result is 4 times smaller than  that obtained by the
MACHO  team.    Consequently,  further  discussion   has  been  opened
concerning  the location  and  nature of  the  lenses. In  particular,
recent LMC  models have  been used in  order to ascertain  if possible
asymmetries  in the space  distribution of  the microlensed  stars are
assimilable to Galactic halo objects  or LMC ones.  In fact, different
studies show  that a sizeable  fraction of the microlensed  events are
due  to LMC  self--lensing (Sahu  1994;  Gyuk, Dalal  \& Griest  2000;
Calchi Novati  et al.   2006).  Moreover, a  full variety  of possible
explanations have been proposed  to reproduce the microlensing events:
tidal debris or a dwarf galaxy  toward the LMC (Zhao 1998), a Galactic
extended  shroud population  of  white dwarfs  (Gates  \& Gyuk  2001),
blending  effects (Belokurov, Evans  \& Le  Du 2003,  2004), spatially
varying mass functions (Kerins \&  Evans 1998, Rahvar 2005), and other
explanations  (Holopainen  et  al.   2006).   However,  all  of  these
proposals have been received with  some criticism because none of them
fully explains the observed microlensing results.

Other  observational  pieces of  evidence  are  added  to conform  the
present  puzzle, such  as the  search for  very faint  objects  in the
Hubble Deep  Field or the  search for the microlensing  events towards
the  Galactic   bulge  or  towards  very  crowded   fields  like  M31.
Particularly,  the   Hubble  Deep  Field  has  provided   us  with  an
opportunity to test  the contribution of white dwarfs  to the Galactic
dark  matter.  Ibata  et  al.   (1999) and,  most  recently, Kilic  et
al. (2005) have  claimed the detection of some  white dwarf candidates
among several  faint blue  objects. These objects  exhibit significant
proper motion  and, thus, are assumed  to belong to the  thick disk or
the  halo  populations.   Despite  the increasing  number  of  surveys
searching  for white  dwarfs ---  like  the Sloan  Digital Sky  Survey
(Eisenstein  et al.   2006), the  2 Micron  All Sky  Survey  (Cutri et
al. 2003),  the SuperCosmos Sky Survey (Hambly,  Irwin \& MacGillivray
2001), the  2dF QSO Redshift Survey  (Vennes et al.   2002), and other
observational  searches  (Knox  et  al.   1999; Ibata  et  al.   1999;
Oppenheimer et al. 2001; Majewski \& Siegel 2002; Nelson et al.  2002)
--- and the  numerous theoretical studies (Isern et  al. 1998; Reyl\'e
et   al.  2001;   Flynn  et   al.  2003;   Hansen  \&   Liebert  2003;
Garc\'{\i}a-Berro et  al.  2004), the  problem still remains  open. In
addition, despite  the fact that, for obvious  reasons, no information
about the halo  could be derived from the  microlensing events towards
the  Galactic  bulge,  the   results  obtained  so  far  indicate  the
primordial role played by low--mass stars rather than other objects.

In two previous papers we have extensively analyzed the role played by
the carbon--oxygen (CO)  white dwarf population (Garc\'{\i}a--Berro et
al.  2004)  as well  as the contribution  of oxygen--neon  (ONe) white
dwarfs (Camacho  et al.  2007).  In  these papers we  have performed a
thorough study of a wide range of Galactic inputs, including different
initial  mass functions and  halo ages,  and several  density profiles
corresponding to  different halo models. The  calculations reported in
these two papers have shown that  a sizeable fraction of the halo dark
matter cannot be locked in the form of white dwarfs.  Specifically, we
have found that this contribution is of the order of a modest $5\%$ in
the most optimistic case, that it is mainly due to old CO white dwarfs
with hydrogen--rich atmospheres and that the contribution of ONe white
dwarfs is  minor, because  although ONe white  dwarfs can  reach faint
magnitudes faster than CO  white dwarfs, their contribution is heavily
suppressed by the initial mass function.

In this paper we analyze in a comprehensive way a wide range of masses
$0.08<  M/M_{\sun}<10$  susceptible  to  produce  microlensing  events
towards the LMC and, thus, to  contribute to the halo dark matter. The
full range of masses studied  here represents nearly the $90\%$ of the
stellar content, including  the red dwarf regime ($M>0.075\,M_{\sun}$)
the CO white dwarf population  and the population of massive ONe white
dwarfs.  The paper  is organized as follows. In  Sect.  2 we summarize
the  main  ingredients  of  our  Monte  Carlo  code  and  other  basic
assumptions  and  procedures necessary  to  evaluate the  microlensing
optical depth  towards the LMC. Section  3 is devoted  to describe our
main  results, including  the  contribution of  red  dwarfs and  white
dwarfs  to the  microlensing optical  depth  towards the  LMC, and  we
compare our  results to those  of the MACHO  and EROS teams.   In this
section  we also estimate  the probability  that a  microlensing event
could be assigned to one or another of the populations under study and
we discuss  the contribution of red  and white dwarfs  to the baryonic
content of  the Galaxy.  Finally, in  Sect.  4 our  major findings are
summarized and we draw our conclusions.

\section{The model}

\subsection{Building the sample}

A detailed description  of our Monte Carlo simulator  has been already
presented in  Torres et al. (2002), Garc\'{\i}a--Berro  et al.  (2004)
and Camacho et al. (2007).  Therefore, we will only summarize here the
most  important  inputs.  The  basic  ingredient of  any  Monte  Carlo
simulator is a  random number generator algorithm which  must ensure a
non--correlated sequence and a  good set of statistical properties. We
have  used a  random  number generator  algorithm  (James 1990)  which
provides a uniform probability density within the interval $(0,1)$ and
ensures a  repetion period of $\ga  10^{18}$, which is  enough for our
purposes.   Each one  of the  Monte Carlo  simulations  presented here
consists of an  ensemble of $\sim5\times10^4$ independent realizations
of  the  synthetic star  population,  for  which  the average  of  any
observational quantity along with its corresponding standard deviation
were computed.  Here the standard deviation means the ensemble mean of
the sample dispersions for a typical sample.

The galactic  halo has been modelled assuming  a spherically symmetric
halo.  In  particular the  model used here  is the  typical isothermal
sphere of radius $5$ kpc  also called the ``S--model'', which has been
extensively  used by  the MACHO  collaboration (Alcock  et  al.  2000;
Griest 1991).  The position of  each synthetic star is randomly chosen
according to this  density profile.  We have not  used other profiles,
such as the exponential power--law model, the Navarro, French \& White
(1997) density  profile  and others  because  their  inclusion in  the
analysis presented here does not provide significant variations in the
final  results  (Garc\'{\i}a--Berro   et  al.  2004).  Likewise,  this
reasoning  is extensible  to the  non--standard galactic  halo models,
such as flattened profiles, oblate  halo models, and others, which are
beyond the scope of the present study.

The mass  distribution of synthetic stars has  been computed according
to  two different initial  mass functions,  the standard  initial mass
function  of Scalo  (1998)  and the  biased  log--normal initial  mass
function  proposed  by  Adams  \&  Laughlin (1996),  the  later  being
representative of other  non--conventional initial mass functions such
as  the one  of Chabrier  et al.   (1996). Note,  however,  that these
biased  initial  mass  functions  seem  to be  incompatible  with  the
observed properties of  the halo white dwarf population  (Isern et al.
1998;  Garc\'\i  a--Berro et  al.   2004),  with  the contribution  of
thermonuclear  supernovae  to the  metallicity  of  the Galactic  halo
(Canal et al.   1997), and with the observations  of galactic halos in
deep galaxy surveys (Charlot \& Silk 1995). Nevertheless, for the sake
of  completeness, we  prefer  to  include in  the  present analysis  a
representative  example of  these  biased mass  function  in order  to
illustrate the role  played by the red dwarf  population when a biased
initial mass function is adopted.

The main sequence mass is  obtained by drawing a pseudo--random number
according to  the adopted initial mass  function. The time  at which a
star was born  is randomly choosen, previously assuming  that the halo
was  formed 14  Gyr  ago in  an  intense burst  of  star formation  of
duration $\sim  1$~Gyr. The main--sequence  lifetime as a  function of
the mass in the main squence is that of Iben \& Laughlin (1989).  Once
the mass of a synthetic star is chosen, the main--sequence lifetime is
obtained and from  it we know which stars are able  to evolve to white
dwarfs or  which ones remain in  the main sequence as  red dwarfs.  We
have   considered   red  dwarfs   to   have   masses   in  the   range
$0.08<M/M_{\sun}<1$. For these stars  we have adopted the evolutionary
models of  Baraffe et  al. (1998). Stars  with such small  masses have
very    large     main--sequence    lifetimes.     Consequently,    no
post--main--sequence   evolutionary  tracks   of   these  stars   were
needed. For  those stars which  have had time  enough to enter  in the
white  dwarf cooling  track, and  given a  set of  theoretical cooling
sequences and the initial to final mass relationship (Iben \& Laughlin
1989) their  luminosities,  effective  temperatures  and  colors  were
obtained. The cooling sequences adopted here depend on the mass of the
white dwarf.  White dwarfs  with masses smaller than $M_{\rm WD}=1.1\,
M_{\sun}$ are expected  to have a CO core  and, consequently, we adopt
for them the  cooling tracks of Salaris et  al.  (2000).  White dwarfs
with masses larger than $M_{\rm WD}=1.1\, M_{\sun}$ most probably have
ONe cores and for these white  dwarfs we adopt the most recent cooling
sequences of Althaus  et al.  (2007).  Both sets  of cooling sequences
incorporate the most accurate physical inputs for the stellar interior
(including  neutrinos,  crystallization,  phase separation  and  Debye
cooling)  and reproduce  the  blue turn  at  low luminosities  (Hansen
1998).

The kinematical  properties of the  halo population have  been modeled
according to gaussian  laws (Binney \& Tremaine 1987)  with radial and
tangential  velocity  dispersions accordingly  related  by the  Jean's
equation and  fulfilling the  flat rotation curve  of our  Galaxy.  We
have  adopted  standard  values  for  the  circular  velocity  $V_{\rm
c}=220$~km/s  as  well  as  for  the  peculiar  velocity  of  the  Sun
$(U_{\sun},  V_{\sun},W_{\sun})=(10.0,  15.0,  8.0)$~km/s  (Dehnen  \&
Binney  1998).   Additionally,  we  have discarded  those  stars  with
velocities smaller than $250$~km/s  given that they are not considered
as  halo members.   Besides  we have  rejected  stars with  velocities
larger than 750~km/s, because they would have velocities exceeding 1.5
times the escape velocity.  Finally, since white dwarfs usually do not
have  determinations of  the  radial component  of  the velocity,  the
radial velocity is eliminated when a comparison with the observational
data is needed.

Finally,  and in  order  to  compare the  simulated  results with  the
observational ones, a normalization criterion should be used.  We have
proceeded  as   in  our  previous   papers  (Camacho  et   al.   2007;
Garc\'{\i}a--Berro  et al.   2004). That  is, we  have  normalized our
simulations to  the local density  of halo white dwarfs  obtained from
the  halo white  dwarf luminosity  function of  Torres et  al.  (1998)
taking into  account the new halo  white dwarf candidates  in the SDSS
Stripe 82 (Vidrih et al.  2007).

\subsection{The LMC microlensing model}

In order to mimic the microlensing experiments towards the LMC we have
simulated it  following closely the detailed LMC  descriptions of Gyuk
et al.  (2000)  and Kallivayalil et al. (2006).   Our model takes into
account, among other parameters, the  scale length and scale height of
the LMC,  its inclination and its kinematical  properties.  This model
provides us with a synthetic population of stars representative of the
monitored point  sources.  Afterwards we have evaluated  which star of
the galactic  halo could be  responsible of a microlensing  event.  We
have only considered stars fulfilling a series of conditions. First of
all the lensing star should be fainter than a certain magnitude limit.
In a  second step we have checked  if the lens is  inside the Einstein
tube of the  monitored star. That is, if  the angular distance between
the lens  and monitored star is  smaller than the  Einstein radius. We
recall here that the Einstein radius is given by

\begin{equation}
R_{\rm E}=2\sqrt{\frac{GMD_{\rm OS}}{c^2}x(1-x)}
\end{equation}

\noindent  where $D_{\rm  OS}$  is the  observer--source distance  and
$x\equiv D_{\rm OL}/D_{\rm OS}$.  Finally, we filter those stars which
are  candidates to  produce a  microlensing event  with  the detection
efficiency function, $\varepsilon(\hat t_i)$,  where $\hat t_i$ is the
Einstein ring diameter crossing time. The detection efficiency depends
on the particular  characteristics of the experiment.  In  our case we
have reproduced the MACHO and EROS experiments.  Specifically, for the
MACHO collaboration we have taken $1.1\times 10^7$ stars during 5.7 yr
and over  $13.4\ {\rm
 deg^2}$, whereas the  detection efficiency has
been modelled according to:

\begin{equation}
\varepsilon(\hat{t})= 
\left\{
\begin{array}{cc}
0.43\,{\rm e}^{-(\ln(\hat{t}/T_{\rm m}))^{3.58}/0.87}, &\hat{t}>T_{\rm m} \\   
43\,{\rm e}^{-|\ln(\hat{t}/T_{\rm m})|^{2.34}/11.16},  &\hat{t}<T_{\rm m}
\end{array}
\right.
\end{equation}

\noindent where $T_{\rm m}=250$ days.  This expression provides a good
fit to the results of Alcock  et al.  (2000).  For the EROS experiment
we  have  used $0.7\times  10^7$  stars over  a  wider  field of  $84\
{\deg^2}$  and over  a period  of $6.7$  yr.  Regarding  the detection
efficiency we have adopted a numerical fit to the results of Tisserand
et al. (2007).

For  all the  simulations  presented here  we  extract the  parameters
relevant  to characterize  the microlensing  experiments.   A complete
description  of  the  various  parameters  which  have  importance  in
discussing  gravitational microlensing  can be  found in  Mollerach \&
Roulet (2002)  and Schneider et  al.  (2004).  Among  these parameters
perhaps the most important one  for our purposes is the optical depth,
$\tau$, which measures the probability that  at a given time a star is
magnified  by  a  lens  by  more  than a  factor  of  1.34.   From  an
observational  point of  view an  estimate  of this  parameter can  be
obtained using the expression (Alcock et al. 2000):

\begin{equation}
\tau=\frac{1}{E}\frac{\pi}{4}\sum_i 
\frac{\hat t_i}{\varepsilon(\hat t_i)}
\end{equation}

\noindent where $E$ is the  total exposure in star--years. The optical
depth  is  independent  of  the  lens motion  and  mass  distribution.
However, since the experiments measure  the number of events and their
durations,   additional  information   can  be   obtained   using  the
microlensing rate,  $\Gamma$, and its distribution as  function of the
event durations.  This parameter  represents nothing else but the flux
of lenses  inside the microlensing  tube. Finally, an estimate  of the
expected number of events can be done using the expression

\begin{equation}
N_{\rm exp}=E\int_{0}^{\infty}\frac{d\Gamma}{d\hat{t}}\varepsilon(\hat
t_i)d\hat t_i
\end{equation}

\section{Results}

\begin{table*}
\centering
\begin{tabular}{lrrrrrrrr}
\hline 
\hline 
\multicolumn{1}{c}{\  } & 
\multicolumn{4}{c}{Standard} &
\multicolumn{4}{c}{AL} \\ 
\hline 
Magnitude                                               & 17.5    & 22.5    & 27.5    & 32.5    & 17.5    & 22.5    & 27.5    & 32.5     \\ 
\cline{2-5} 
\cline{6-9} 
$\langle N_{\rm WD}\rangle$                             & $0\pm1$ & $0\pm1$ & $0\pm1$ & $0\pm1$ & $3\pm2$ & $2\pm1$ & $1\pm1$ & $0\pm1$  \\ 
$\langle m \rangle$  $(M/M_{\sun})$                     & 0.421   & 0.411   & 0.427   & 0.443   & 0.638   & 0.636   & 0.640   & 0.684    \\  
$\langle\mu\rangle$  $(^{\prime\prime}\,{\rm yr}^{-1})$ & 0.020   & 0.017   & 0.008   & 0.004   & 0.036   & 0.025   & 0.011   & 0.003    \\ 
$\langle d\rangle$ (kpc)                                & 2.48    & 3.79    & 6.62    & 13.08   & 1.39    & 2.15    & 5.13    & 19.6     \\  
$\langle V_{\rm  tan}\rangle$ $({\rm km\,s}^{-1})$      & 240     & 247     & 262     & 241     & 240     & 252     & 263     & 261      \\ 
$\langle\hat{t}_{\rm E}\rangle $ (d)                    & 41.2    & 49.3    & 63.3    & 82.8    & 34.7    & 46.6    & 76.4    & 126.8    \\  
$\langle \tau/\tau_0  \rangle$                          & 0.283   & 0.214   & 0.139   & 0.055   & 0.302   & 0.204   & 0.140   & 0.129    \\ 
\hline
\hline
\end{tabular}
\caption{Summary  of  the results  obtained  for  the whole  simulated
        population of  microlenses towards the  LMC for an age  of the
        halo of  14~Gyr, different  model initial  mass functions, and  
        several magnitude cuts.}
\end{table*}

\subsection{Optical depth towards the LMC}

As previously mentioned,  the optical depth provides us  with the most
immediate and simplest information about the microlensing experiments.
Even so, the optical depth plays  a key point in our analysis since it
provides us with rich information about the Galactic halo and the role
of the dark  matter.  We will compare here  the results obtained using
our  Monte   Carlo  simulator  with   those  obtained  by   the  MACHO
collaboration.  In  Fig.  1  we show the  contribution to  the optical
depth of the different simulated  populations for the two initial mass
functions under study as a function of the adopted magnitude cut.  Our
simulations have been normalized to the value derived by Alcock et al.
(2000), $\tau_0=1.2\times  10^{-7}$.  The white  dwarf populations are
represented by solid and open squares for the CO and ONe white dwarfs,
respectively.   Open triangles  show the  contribution of  red dwarfs,
while the contribution of the three populations is represented by open
circles.  The first remarkable result is that for the standard initial
mass function (top panel of  Fig.  1) the combined contribution of red
dwarfs and white  dwarfs is at most one third  of the observed optical
depth and this is obtained when a totally unrealistic magnitude cut is
adopted, $m_V\sim 15^{\rm mag}$.  Note as well that there is a clearly
decreasing trend as the adopted  magnitude cut increases.  This can be
easily understood, since fewer objects contribute to the optical depth
as the  magnitude cut increases.  Furthermore, the  slope is different
for  the three  types of  objects considered  here. For  instance, the
contribution of  red dwarfs decreases faster than  the contribution of
CO  white  dwarfs as  the  magnitude  cut  increases which,  in  turn,
decreases  faster than  the contribution  of ONe  white  dwarfs.  This
reflects  the fact  that,  in  general, red  dwarfs  are bighter  than
regular CO  white dwarfs.   Finally, ONe white  dwarfs cool  very fast
(Althaus et  al.  2007) and thus,  for realistic halo  ages, are faint
objects.   Consequently, their  contribution remains  almost constant.
It is interesting to note as  well that the optical depth obtained for
the whole population almost doubles  that obtained when only the white
dwarf population  is taken into  account.  However, in  agreement with
other studies,  the value  obtained here still  remains far  below the
observed one and it is thus clear that the Galactic halo population is
not  enough  to explain  the  results of  the  MACHO  team.  Even  the
alternative initial mass functions, of which that of Adams \& Laughlin
(1996) is   a   representative    example,   predict   that   $\langle
\tau/\tau_0\rangle$ is, in the best of the cases, 0.3 and this happens
for unrealistic  magnitude cuts  --- see the  bottom panel of  Fig. 1.
The  results obtained  using the  non--standard initial  mass function
deserve two additional comments.   In particular, it is interesting to
note that  the largest  contribution in  this case is  that of  the CO
white  dwarf population and,  secondly, that  the contribution  of red
dwarfs is  totally negligible for magnitude cuts  larger than $m_V\sim
23^{\rm mag}$, which is a  reasonable value for current surveys.  Both
facts are natural  since this initial mass function  has been tailored
to produce  a large population  of $0.5\, M_{\sun}$ white  dwarfs.  In
summary,  the  fraction of  the  microlensing  optical  depth which  a
stellar halo can  account for in the range of  masses under study here
is at most one third of that observed by the MACHO team, independently
of the adopted initial mass function.

\begin{figure}
\vspace{13cm}    
\includegraphics{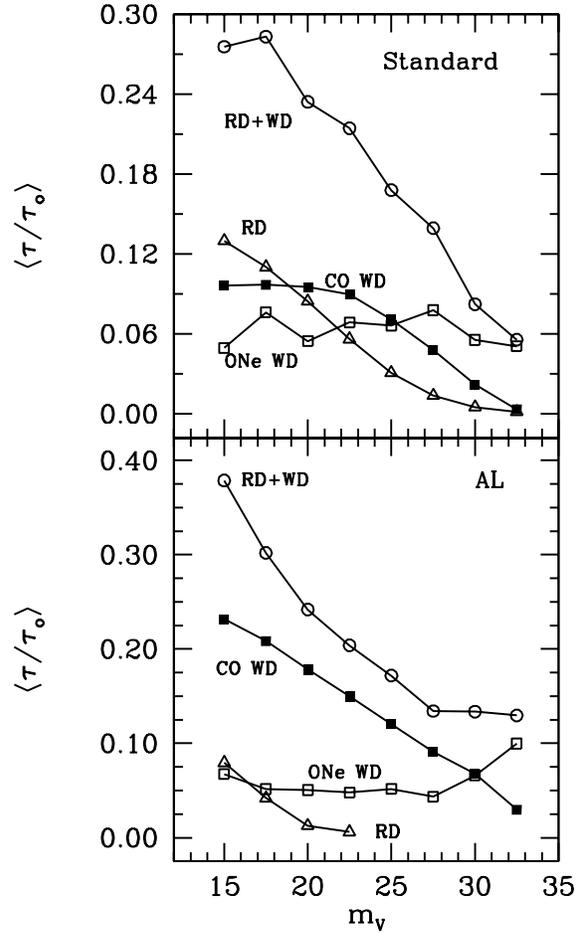} 
\caption{Microlensing optical  depth towards the LMC as  a function of
         the limiting magnitude. Solid  and open squares represent the
         CO and ONe white  dwarf populations, respectively. Red dwarfs
         are  represented  using   open  triangles,  while  the  whole
         population is shown using open circles.}
\end{figure}

In Table  1 we  have summarized the  average values for  the different
parameters of the whole population  for the two initial mass functions
under study and  for four magnitude cuts. Specifically,  in Table 1 we
present the  number of  microlensing events, the  average mass  of the
microlenses,  their  average proper  motion,  distance and  tangential
velocity, the corresponding Einstein  crossing times and, finally, the
contribution to the microlensing optical depth.  It is clear that some
of these parameters are to some extent dependent of the magnitude cut.
For  instance, the  average distance  of the  sample increases  as the
magnitude cut  is larger. This  is a natural consequence  of selecting
more distant objects which,  in turn, implies longer Einstein crossing
times.  This  behavior is  independent  of  the  assumed initial  mass
function. However,  there are characteristics  which do not  change as
the magnitude  cut increases.  For instance, this  is the case  of the
expected number of  events or the average mass of  the sample.  In the
case of a standard initial mass function no more than one microlensing
event should be expected at  the $1\sigma$ confidence level, while for
a log--normal mass function up to  5 events might be expected.  In any
case the  expected number  of microlensing events  is far from  the 17
events claimed by the MACHO experiment.

As already mentioned,  the average mass of the  microlenses depends on
the initial mass function.  To  investigate this further in Fig. 2 the
contribution to  the optical depth  as a function  of the mass  of the
lens  object for  both initial  mass functions  is shown.  The results
obtained  with our  Monte  Carlo  simulator clearly  show  that for  a
standard initial mass function --- top  panel of Fig.  2 --- there are
two  peaks   centered  at   masses  $\sim  0.3\,M_{\sun}$   and  $\sim
0.6\,M_{\sun}$, respectively.  These  masses correspond to the average
masses  of  red  dwarfs  and  CO  white  dwarfs,  respectively.   Also
noticeable is  that the contribution of  ONe white dwarfs  can only be
appreciated as an extended tail in the case of a standard initial mass
function.  This is  in full  agreement with  recent studies  about the
distribution of  masses of the  white dwarf population (Finley  et al.
1997;  Liebert et al.   2005), which  show the  existence of  a narrow
sharp peak near $0.6\,M_{\sun}$,  with a tail extending towards larger
masses,  with several white  dwarfs with  spectroscopically determined
masses   within    the   interval   comprised    between   $1.0$   and
$1.2\,M_{\sun}$.   The situation  is different  for  the non--standard
initial  mass  function,  which  is  shown  in  the  bottom  panel  of
Fig. 2.  The log--normal initial mass function  considered here cannot
produce red  dwarfs with masses  below $\sim 0.45\,M_{\sun}$  and thus
the peak at $0.3\, M_{\sun}$ previously found is absent in this case.

\begin{figure}
\vspace{13cm}
\includegraphics{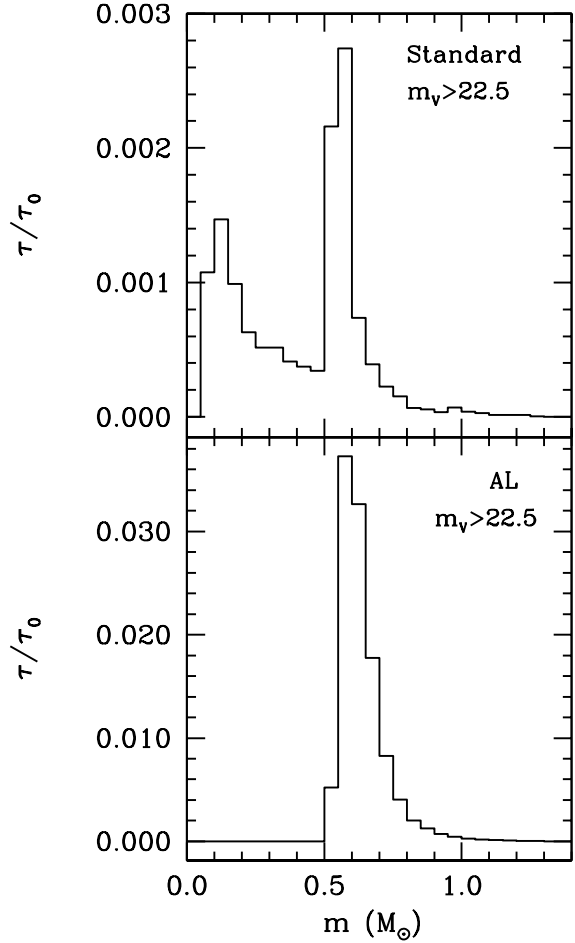}
\caption{Contribution to the  optical depth as a function  of the lens
        mass.}
\end{figure}

\begin{table*}
\centering
\begin{tabular}{lrrrrrrrr}
\hline
\hline
\multicolumn{1}{c}{\ } &
\multicolumn{4}{c}{Standard} &
\multicolumn{4}{c}{AL}  \\
\hline
Magnitude                                               & 17.5     & 22.5     &  27.5    & 32.5     & 17.5      & 22.5     & 27.5     & 32.5      \\
\cline{2-5}
\cline{6-9}
$\langle N_{\rm WD}\rangle$                             & $0\pm 1$ & $0\pm 1$ & $0\pm 1$ & $0\pm 1$ & $0\pm  1$ & $0\pm 1$ & $0\pm 0$ & $0\pm  0$ \\  
$\langle m \rangle$ $(M/M_{\sun})$                      & 0.324    & 0.228    & 0.119    & 0.092    & 0.747     & 0.622    & ---      & ---       \\ 
$\langle \mu\rangle$ $(^{\prime\prime}\,{\rm yr}^{-1})$ & 0.018    & 0.011    & 0.006    & 0.005    & 0.010     & 0.002    & ---      & ---       \\
$\langle d\rangle$ (kpc)                                & 2.89     & 4.88     & 8.27     & 10.20    & 5.15      & 17.59    & ---      & ---       \\ 
$\langle V_{\rm tan}\rangle$ $({\rm km\,s}^{-1})$       & 242      & 244      & 254      & 235      & 246       & 180      & ---      & ---       \\
$\langle\hat{t}_{\rm E}\rangle  $ (d)                   & 41.2     & 48.5     & 41.3     & 43.3     & 98.2      & 158.0    & ---      & ---       \\
$\langle \tau/\tau_0\rangle$                            & 0.130    & 0.118    & 0.096    & 0.070    & 0.091     & 0.125    & ---      & ---       \\
\hline
\hline
\end{tabular}
\caption{Summary  of the results  obtained for  the population  of red
	 dwarf microlenses towards  the LMC for an age  of the halo of
	 14~Gyr,  different  model initial mass functions, and several 
         magnitude cuts.}
\end{table*}

A  more  detailed  analysis  of  the  role played  by  the  red  dwarf
population can be done. To this  end in Table 2 we have summarized the
average values for  the different parameters of the  population of red
dwarfs for both initial mass  functions.  Similar sets of data for the
white dwarf population  can be found in our  previous studies (Camacho
et al.  2007,  Garc\'{\i}a--Berro et al.  2004).  As  can be seen from
Table 2,  the red dwarf population  roughly contributes a  10\% to the
observed MACHO optical depth for a standard initial mass function.  It
is also  important to discuss the  other parameters shown  in Table 2.
For instance,  the average mass  clearly decreases when  the magnitude
cut  increases, which is  the opposite  to what  occurs for  the white
dwarf population.  This can be easily explained by taking into account
that the more massive the red  dwarf is the brighter is and, hence, we
should expect  less massive objects  for larger magnitude  cuts.  This
result is reinforced  by the fact that the  average distance increases
for increasing magnitude cuts.  Moreover, since the average tangential
velocity  remains   constant,  the  combined  effect   of  an  average
decreasing  mass  and  an  average  increasing distance  is  that  the
Einstein   crossing   time    remains   practically   constant.    The
characteristics of the red dwarf population are dramatically different
when the initial mass function of Adams \& Laughlin (1996) is used. In
this  case   the  production  of  low--mass  red   dwarfs  is  heavily
suppressed.  Accordingly, in our  simulations we have not produced red
dwarfs with masses smaller  than $\sim 0.45\,M_{\sun}$. Thus, since in
average the masses are larger  we also find brighter stars.  Hence, we
expect  no contribution  at all  for magnitude  cuts  above $m_V\simeq
22.5^{\rm mag}$,  while for brighter  magnitude cuts the  average mass
expected  is  $\sim 0.7\,M_{\sun}$,  which  is  even  larger than  the
expected value for CO white dwarfs.

As  already   seen,  for  the   standard  initial  mass   function,  a
double--peaked  profile is found,  but the  peak amplitude  deserves a
detailed analysis. The ratio of  the contribution to the optical depth
of a typical  red dwarf with respect to the  contribution of a typical
CO white dwarf is

\begin{figure}
\vspace{13cm}
\includegraphics{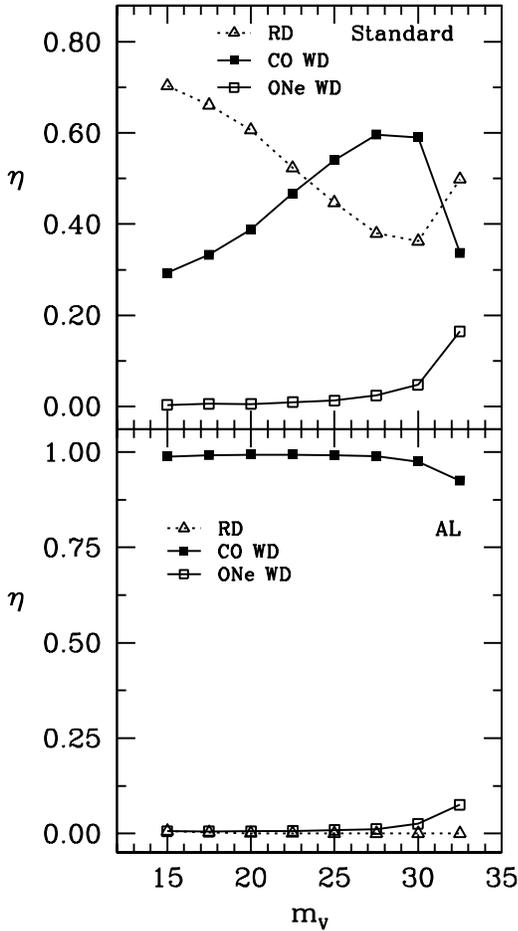}
\caption{Fraction   of   microlenses  with   respect   to  the   whole
         population, as a function of the magnitude cut.}
\end{figure}

\begin{equation}
\frac{\tau_{\rm RD}}{\tau_{\rm CO}}= 
\frac{\hat t_{\rm RD}}{\hat t_{\rm  CO}} 
\frac{\varepsilon(\hat t_{\rm CO})}{\varepsilon(\hat t_{\rm  RD})}
\approx\sqrt{
\frac{M_{\rm RD}D_{\rm OL}^{\rm  RD}}{M_{\rm CO}{D_{\rm OL}^{\rm CO}}}}
\frac{\varepsilon(\hat t_{\rm CO})}{\varepsilon(\hat t_{\rm RD})}
\end{equation}

\noindent This value depends on  the adopted initial mass function and
on the magnitude cut used.  For the standard initial mass function and
a magnitude cut of $22.5^{\rm mag}$ the average values of the mass and
distance of a red  dwarf are, respectively, $\sim 0.228\,M_{\sun}$ and
$4.88\,$kpc, while  for a typical CO  white dwarf the  average mass is
$\sim 0.568\,M_{\sun}$ and the  average distance is $3.14\,$kpc.  With
these  values  the  ratio  of  the  optical depths  turns  out  to  be
$\tau_{\rm RD}/\tau_{\rm  CO}\approx 0.9$.  Thus,  although the sample
space of red dwarfs and  white dwarfs is different the contribution to
the optical depth per object results to be the same. For instance, the
red dwarf population has a smaller average mass than white dwarfs but,
conversely,  the  average distance  is  larger.  Therefore, the  total
contribution of these two populations  results to be merely an account
of the number  of microlenses. In order to clarify  this issue we have
evaluated the fraction of microlenses due to the different populations
as a function  of the adopted magnitude cut. The  results are shown in
Fig. 3 for  both the standard initial mass function  --- top panel ---
and the initial  mass function of Adams \&  Laughlin (1996) --- bottom
panel.   As  it can  be  seen there,  for  the  standard initial  mass
function  the  relative  contribution  of red  dwarfs  decreases  with
increasing magnitude  cuts, while that  of CO white  dwarfs increases.
Both contributions are  equal for a magnitude cut  of $\approx 24^{\rm
mag}$.  Finally, the contribution  of ONe white dwarfs remains roughly
constant and  only becomes appreciable for very  large magnitude cuts.
These trends  can be attributed to  the fact that red  dwarfs are more
numerous  at bright magnitudes  than white  dwarfs, for  which typical
luminosities  are of the  order of  $\log(L/L_{\sun})\simeq-3.5$.  The
situation is  completely different  when the log--normal  initial mass
function of Adams  \& Laughlin (1996) is used.  As can  be seen in the
bottom  panel of  Fig. 3,  the  number of  microlenses is  practically
dominated by  the CO white dwarf contribution,  while the contribution
of red dwarfs and ONe white dwarfs is negligible.

\subsection{The microlensing event rate}

As previously pointed out, the contribution to the optical depth for a
standard  initial  mass  function   is  doubled  when  the  red  dwarf
population  is considered.   We have  also shown  that for  a standard
initial mass  function the  contributions of red  dwarfs and  CO white
dwarfs are roughly the same.  Then,  it is natural to ask ourselves if
there are differences which can help us in discerning the contribution
of one or another population using the observational data of the MACHO
experiment. To assess this we have analyzed the microlensing rate as a
function of the event duration.   The results of our simulations for a
standard  initial mass function  are shown  in Fig.   4.  Each  of the
panels  is clearly  labelled with  the adopted  magnitude cut  and the
population of microlenses.  For the case of the left panels in Fig.  4
we  have adopted a  magnitude cut  of $25^{\rm  mag}$, whilst  for the
right panels  a magnitude cut of  $30^{\rm mag}$ was  adopted.  In all
the cases the simulated microlensing  rate is shown using solid lines,
while the observational data obtained by the MACHO team is shown using
a dotted line. All the  distributions are normalized to unit area.  As
can be seen,  the red dwarf and the  white dwarf distributions present
some differences.  For instance, for a magnitude cut of $25^{\rm mag}$
it turns out that although both the red dwarf population and the white
dwarf populations show a peak  located at nearly $\sim 20\,$ days, the
white dwarf population  presents a wider distribution. In  the case in
which a magnitude cut of $30^{\rm mag}$ is adopted the differences are
more pronounced and it is clear  that the peak of the distribution for
the CO white population moves  to longer durations ($\sim 70\,$ days),
whereas  the  peak   of  the  red  dwarf  population   does  not  move
appreciably.

\begin{figure*}
\vspace{12.5cm}
\includegraphics{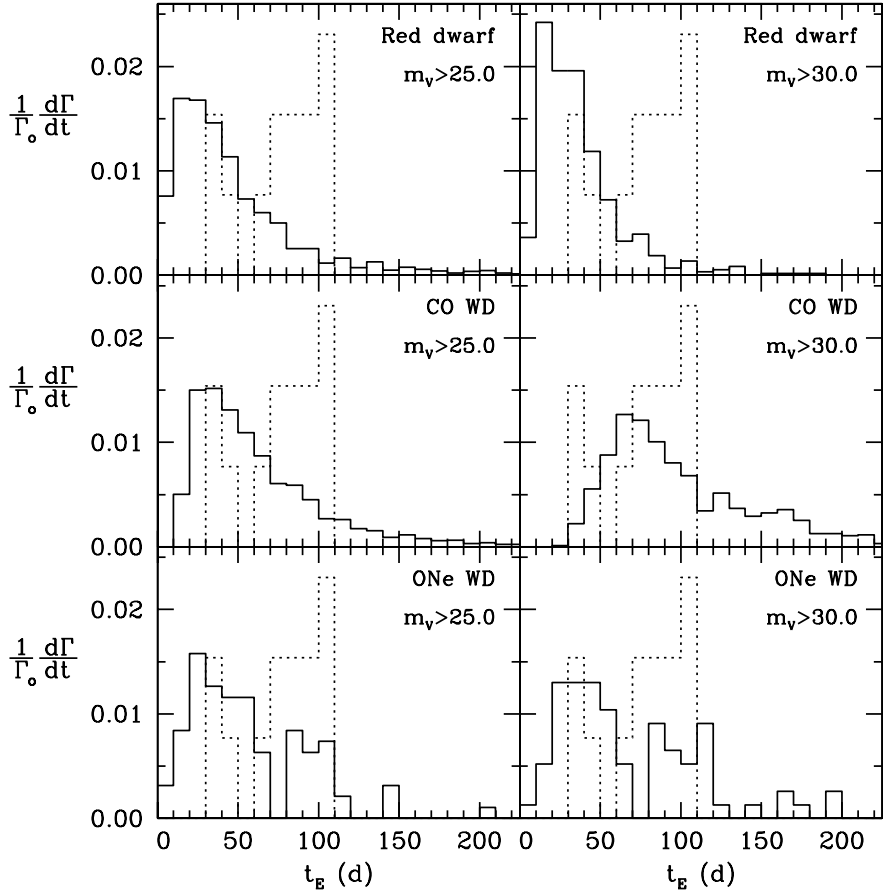}
\caption{Differential event rate normalized to unit area as a function
         of the Einstein crossing time for the populations under study
         and different magnitude  cuts (solid lines). Also represented
         as a  dotted line  in each panel  is the  observational event
         distribution from Alcock et al. (2000). }
\end{figure*}

\begin{table*}
\centering { $Z ^2$ COMPATIBILITY TEST}\\
\vspace{0.2cm}
\begin{tabular}{lcccccc}
\hline 
\hline  
Magnitude        & $ 17.5^{\rm  mag}$ & $ 22.5^{\rm  mag}$ & $ 25.0^{\rm mag}$ & $ 27.5^{\rm mag}$ & $ 30.0^{\rm mag}$ \\ 
\cline{2-6}
Red dwarfs       & 0.58             & 0.67               & 0.65                & 0.55              & 0.40              \\
CO white dwarfs  & 0.61             & 0.70               & 0.80                & 0.87              & 0.90              \\
ONe white dwarfs & 0.76             & 0.55               & 0.68                & 0.68              & 0.72              \\
Whole population & 0.59             & 0.69               & 0.75                & 0.82              & 0.89              \\
\hline
\hline
\end{tabular}
\caption{Compatibility, as obtained  using the $Z^2$ statistical test,
         of the observed MACHO distribution of Einstein crossing times
         and those of the different simulated populations.}
\end{table*}

Additionally,  and in order  to make  more quantitative  estimates, we
have performed  a $Z^2$ statistical  test of the compatibility  of the
different populations  under study with the observed  data.  The $Z^2$
statistical  test  (Lucy 2000)  represents  and  improvement over  the
standard $\chi^2$  statistical test and it is  especially designed for
meagre data  sets. In  Table 3  we show the  $Z^2$ probability  of the
different simulated populations being compatible with the distribution
of  Einstein times  obtained by  the MACHO  experiment.  It  should be
clarified that  this probability represents an estimate  of the degree
to which  the observed event  rate distribution derives from  a single
population of  stars. As can  be seen in  Table 3, the CO  white dwarf
population  best  matches  the  observational  data,  given  that  its
compatibility  is as  high as  0.90  for the  faintest magnitude  cut.
Moreover the  compatibility of this population  with the observational
data  increases for  fainter magnitude  bins. In  sharp  contrast, the
population of red dwarfs presents  a decreasing trend as the magnitude
cut   increases  and,   additionally,  the   compatibility   with  the
observational data is  0.70 in the best of the  cases.  With regard to
the ONe  white dwarf population the compatibility  presents and almost
constant  value around  $0.70$,  independently of  the magnitude  cut.
These  results  indicate  that  the  CO  white  dwarf  population  can
reproduce the observed distribution  of microlensing event rates. Even
more, they  dominate the behavior of  the whole population,  as can be
seen by taking a look at the  last row of Table 3, in which we analyze
the  compatibility  of  the   whole  population  of  simulated  stars.
Therefore, even if the expected number of microlensing events obtained
in our simulations  is considerable smaller than the  $\sim 15$ events
claimed by the MACHO team, the event rate distribution of the CO white
dwarf   population   is   in   fair  agreement   with   the   observed
distribution. This result  puts into question to which  extent we know
the  characteristics of  the halo  white  dwarf population  and if  it
exists other possible ways capable of producing a larger number of old
white dwarfs in the stellar halo.

\begin{table*}
\centering
\begin{tabular}{lrrrrrr}
\hline 
\hline 
\multicolumn{1}{c}{\ } & 
\multicolumn{3}{c}{Standard} &
\multicolumn{3}{c}{AL} \\ 
\hline Magnitude                                       & 17.5      & 22.5     & 27.5      & 17.5      & 22.5      & 27.5      \\ 
\cline{2-4} 
\cline{5-7}
$\langle N_{\rm WD}\rangle$                            & $0 \pm 1$ & $0\pm 1$ & $0 \pm 1$ & $1 \pm 1$ & $1 \pm 1$ & $1 \pm 1$ \\ 
$\langle m \rangle$ $(M/M_{\sun})$                     & 0.385     & 0.384    & 0.427     & 0.633     & 0.637     & 0.637     \\  
$\langle\mu\rangle$ $(^{\prime\prime}\,{\rm yr}^{-1})$ & 0.020     & 0.013    & 0.009     & 0.028     & 0.022     & 0.011     \\ 
$\langle d\rangle$ (kpc)                               & 2.49      & 4.26     & 6.55      & 1.83      & 2.39      & 5.27      \\ 
$\langle V_{\rm tan}\rangle$ $({\rm km\,s}^{-1})$      & 241       & 269      & 267       & 242       & 250       & 266       \\ 
$\langle \hat{t}_{\rm E}\rangle$ (d)                   & 38.3      & 45.0     & 54.7      & 42.6      & 50.0      & 75.6      \\  
$\langle\tau/\tau_0\rangle$                            & 0.558     & 0.695    & 0.163     & 0.839     & 0.628     & 0.488     \\ 
\hline 
\hline
\end{tabular}
\caption{Summary  of  the  results  obtained  for  the  simulation  of
	microlenses towards the LMC for the EROS experiment for an age
	of the halo of 14~Gyr, different model initial mass functions, 
        and  several magnitude cuts.}
\end{table*}

\subsection{The EROS experiment}

While the MACHO  team claims for the identification  up to 17 observed
events, the  EROS collaboration have not found  any microlensing event
towards the  LMC and one candidate  event towards the  SMC. Adopting a
standard halo  model and assuming  $\tau_{\rm SMC}=1.4\tau_{\rm LMC}$,
the  EROS results  imply an  optical  depth $\tau_0=0.36\times10^{-7}$
(Tisserand  et al.   2007),  which  is four  times  smaller than  that
obtained  by  the  MACHO  team.   We  have also  performed  a  set  of
simulations emulating  the conditions of the EROS  experiment with the
same  inputs previously  described in  Sect. 2.   Although  only minor
differences should be expected in the analysis of the main results, it
is  clear as  well that  a  joint study  of both  experiments using  a
controlled set of prescriptions represents a test of the robustness of
our numerical procedure.

In Table 4 we have summarized  the results obtained in this second set
of Monte Carlo simulations of microlenses towards the LMC for the EROS
experiment.  Our simulations show  that, for the standard initial mass
function,  the expected  optical depth  could be  $70\%$ of  the value
found by the  EROS team. The value obtained when  only the white dwarf
population was considered has been  found to be $50\%$ (Camacho et al.
2007).   Thus, the  simulations  presented here  reproduce better  the
results of the EROS experiment.  Obviously the red dwarf population is
the  responsible  of  this  increment.   On the  other  hand,  when  a
non--standard initial mass function  is adopted, the results show only
marginal differences with respect to  those obtained for a white dwarf
population,  given  that  in this  case  the  role  of red  dwarfs  is
extremely limited. In summary, our  results are in fair agreement with
those obtained by  the EROS experiment,and they seem  to indicate that
the microlensing optical depth  obtained by the MACHO collaboration is
an overestimate.

\begin{figure}
\vspace{8cm}
\includegraphics{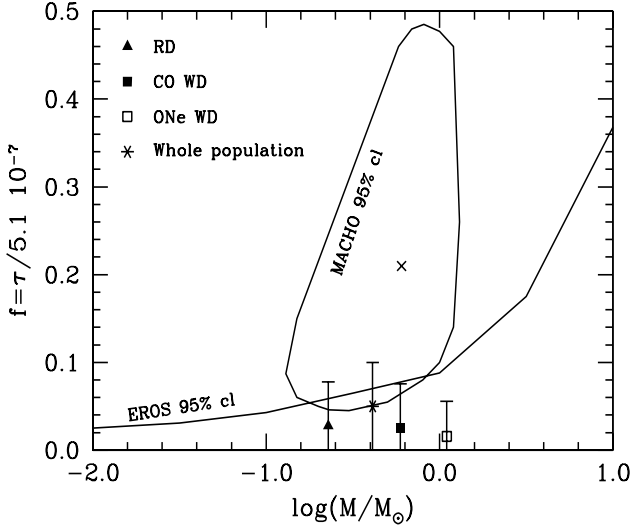} 
\caption{Halo  dark  matter  fraction   as  a  function  of  the  mass
         lens. Plotted  using a solid  line are the  $95\%$ confidence
         level  curve   for  the  MACHO  experiment   and  the  $95\%$
         confidence level upper limit for the EROS experiment.}
\end{figure}

\subsection{The dark matter density}

With the information obtained so far,  we are in position to assess to
which extent the mass range the stellar populations discussed here can
contribute to the baryonic dark  matter halo. Based on their $\sim 15$
microlensing events,  the MACHO collaboration has  derived an estimate
of the halo fraction of dark matter  $f$ as well as the MACHO mass $m$
using maximum--likelihood techniques. A similar analysis has been done
by the  EROS team, but with  the significant difference  that, in this
case, no event has been reported for the LMC, which means that only an
upper limit  on the halo mass  fraction can be obtained.   In order to
compare  the results  of the  MACHO and  EROS collaborations  with our
Monte Carlo  simulations we  have adopted as  our reference  model the
isothermal   sphere  of   core  radius   5   kpc  with   a  value   of
$\rho_0=0.0079\, M_{\sun}$~pc$^{-3}$ for the local dark matter density
and disregarding the contribution of  the LMC halo.  For this model we
obtain  that   the  optical  depth  towards  the   LMC  is  $\tau_{\rm
LMC}=5.1\times  10^{-7}f$. The  different estimates  of the  halo mass
fraction $f$ as a function of the  mass are plotted in Fig. 5. We show
as a  solid line  the curve  of the MACHO  $95\%$ confidence  level as
taken from Alcock et al. (2000)  as well as the EROS $95\%$ confidence
level upper limit based on no observed events in the EROS-1 and EROS-2
data (Tisserand  et al.  2007).   Also represented are  the individual
contribution  of each  population under  study  as well  as the  whole
population  along  with their  corresponding  $95\%$ confidence  level
error bars.   It is remarkable that  the value obtained  for the whole
halo simulated population is in agreement within the $95\%$ confidence
level  curves  of both  observational  estimates.   Thus, our  results
predict  that   the  range  of   stellar  masses  within   $0.08$  and
$10\,M_{\sun}$  provides  $f=0.05$ and  an  average  mass of  $0.411\,
M_{\sun}$  to the  halo dark  matter in  agreement with  the currently
observational data.   This result corroborates  our previous estimates
about  the limited contribution  of the  CO white  dwarfs and  the ONe
white dwarfs (Garc\'\i a--Berro et al. 2004; Camacho et al. 2007).

\section{Conclusions}

We have  extended our previous  studies about the contribution  to the
halo  dark  matter of  the  white  dwarf  population to  the  Galactic
population of red dwarfs. Based on a series of Monte Carlo simulations
which incorporate  the most  up--to--date evolutionary tracks  for red
dwarfs, CO white  dwarfs and ONe white dwarfs  and reliable models for
our Galaxy  and the  LMC we have  estimated the contribution  of these
objects to the microlensing optical depth towards the LMC and compared
it with that obtained by the MACHO and EROS collaborations. In a first
set  of simulations we  have found  that the  contribution of  the red
dwarf population practically doubles the contribution found so far for
the  white dwarf  population.   Our results  indicate  that the  whole
population of these stars can account  at most for a $\sim 0.3$ of the
optical depth  found by the MACHO  team.  This value  implies that the
contribution  of the  full range  of masses  between $0.08$  and $10\,
M_{\sun}$ represents a  $5\%$ of the halo dark  matter with an average
mass  of $0.4\,  M_{\sun}$. Despite  that  this result  is in  partial
agreement  with  the $95\%$  confidence  level  MACHO  estimate for  a
standard isothermal sphere and  no halo LMC contribution, the expected
number of events  obtained by our simulations (3  events at the $95\%$
confidence level) is  substantially below the 13 to  17 observed MACHO
events.  These arguments reinforce the idea, previously pointed out by
other studies, that  the optical depth found by  the MACHO team should
be  an overstimate,  probably  due to  contamination of  self--lensing
objects, variable  stars and others.   Moreover, we have  assessed the
compatibility  between the  observed event  rate distribution  and the
ones obtained for the  different populations under study.  Our results
show that the CO white  dwarf population can reproduce fairly well the
observed event  rate distribution although, as  mentioned earlier, the
expected number of events is considerable smaller.  On the other hand,
the negative results obtained by the  EROS team towards the LMC are in
agreement  with our standard  halo simulation.   Finally, and  for the
sake of  completeness, we  have studied the  effects of  a log--normal
biased initial  mass function. In  this case, the contribution  of the
red dwarf  population is  only marginal given  that the  production of
low--mass  stars   is  strongly  inhibited.   Accordingly,  the  total
contribution to  the microlensing optical depth is  not different from
that found in previous studies for the white dwarf contribution.


\begin{acknowledgements}
Part   of    this   work   was    supported   by   the    MEC   grants
AYA05--08013--C03--01 and 02, by the European Union FEDER funds and by
the AGAUR.
\end{acknowledgements}


\end{document}